\documentclass[aps,showpacs,pra,reprint,groupedaddress,amsmath,amssymb,superscriptaddress]{revtex4-2}

\usepackage{amssymb,amsmath}
\usepackage{marvosym}
\usepackage{bm}
\usepackage{hyperref}
\usepackage{graphicx}
\usepackage{xcolor}
\usepackage{mathtools}
\usepackage{makecell,tabularx}
\usepackage{siunitx}
\usepackage[latin1]{inputenc}

\newcommand{\ee}{\mathrm{e}}
\newcommand{\ii}{\mathrm{i}}

\newcommand{\bmkL}{\bm{k}_{\mathrm{L}}}
\newcommand{\kL}{k_{\mathrm{L}}}
\newcommand{\wL}{\omega_{\mathrm{L}}}
\newcommand{\wat}{\omega_{0}}
\newcommand{\dL}{\delta_{\mathrm{L}}}
\newcommand{\OmegaL}{\Omega_{\mathrm{L}}}
\newcommand{\gsp}{\gamma_{\mathrm{sp}}}
\newcommand{\epsilonL}{ \bm{\varepsilon}_{\mathrm{L}} }
\newcommand{\mYb}{m_{\mathrm{Yb}}}
\newcommand{\lambdaLat}{\lambda_{\mathrm{lat}}}
\newcommand{\ER}{E_{\mathrm{R}}}

\begin{document}

\title{Dynamics of spatial phase coherence in a dissipative Bose-Hubbard atomic system}


\author{R\'emy Vatr\'e}

\author{Rapha\"el Bouganne}

\author{Manel Bosch Aguilera}

\author{Alexis Ghermaoui}

\author{J\'er\^ome Beugnon}

\author{Raphael Lopes}

\author{Fabrice Gerbier}
\email[]{fabrice.gerbier@lkb.ens.fr}
\affiliation{Laboratoire Kastler Brossel, Coll\`ège de France, CNRS, ENS-Universit\'e PSL,
Sorbonne Universit\'e, 11 place Marcelin Berthelot, F-75231 Paris, France}


\date{\today}

\begin{abstract}
We investigate the loss of spatial coherence of one-dimensional bosonic gases in optical lattices illuminated by a near-resonant excitation laser. Because the atoms recoil in a random direction after each  spontaneous emission, the atomic momentum distribution progressively broadens. Equivalently, the spatial correlation function (the Fourier-conjugate quantity of the momentum distribution) progressively narrows down as more photons are scattered. Here we measure the correlation function of the matter field for fixed distances corresponding to nearest-neighbor (n-n) and next-nearest-neighbor (n-n-n) sites of the optical lattice as a function of time, hereafter called n-n and n-n-n correlators. For strongly interacting lattice gases, we find that the n-n correlator $C_1$ decays as a power-law at long times, $C_1\propto 1/t^{\alpha}$, in stark contrast with the exponential decay expected for independent particles. The power-law decay reflects a non-trivial dissipative many-body dynamics, where  interactions change drastically the interplay between fluorescence destroying spatial coherence, and coherent tunnelling between neighboring sites restoring spatial coherence at short distances. The observed decay exponent $\alpha \approx 0.54(6) $ is in good agreement with the prediction $\alpha=1/2$ from a dissipative Bose-Hubbard model accounting for the fluorescence-induced decoherence. Furthermore, we find that the n-n correlator $C_1$ controls the n-n-n correlator $C_2$ through the relation $C_2 \approx C_1^2$, also in accordance with the dissipative Bose-Hubbard model.
\end{abstract}

\maketitle

%
%

\section{Introduction}
Interference is a central phenomenon in quantum mechanics\,\cite{schiffQM}. In the early days of its development, the realization that the mechanical properties of a quantum particle can display interference phenomena led to the discovery of wave-particle duality, often presented in introductory textbooks as the first example of the counter-intuitive features of quantum theory.
Today, matter wave interferences are routinely observed experimentally and used in high-precision gravity or rotation sensors. 

The textbook picture of a perfectly coherent matter wave is usually far from experimental realities. In a real system, phase coherence is usually maintained only in a coherence volume $l_{\mathrm{c}}^\mathcal{D}$, with $l_{\mathrm{c}}$ the characteristic \textit{coherence length} and $\mathcal{D}$ the dimensionality. More formally, the degree of spatial phase coherence of a quantum system can be quantified by the first-order correlation function of the quantized matter field $\hat{\Psi}$ (also called single-particle density matrix--SPDM)\,\cite{cct2001a}, 
\begin{align}
\rho^{(1)}(\bm{s})=\int d^{\mathcal{D}}\bm{R}\,\langle \hat{\Psi}^\dagger(\bm{R}+\bm{s}/2) \hat{\Psi}(\bm{R}-\bm{s}/2) \rangle.
\end{align}
Here and in the following, the brackets $\langle \cdot \rangle$ denote the expectation value. Physically, the SPDM $\rho^{(1)}(\bm{s})$ quantifies the contrast of a matter wave interferometer with an arm separation $\bm{s}$. Besides interferometry, the SPDM is also key to the description of degenerate quantum fluids. Long-range spatial phase coherence where $l_{\mathrm{c}}$  is comparable to the size of the system  is  the hallmark of Bose-Einstein condensation in an ultracold gas of bosonic atoms\,\cite{cct2001a,andrews1997a,hagley1999a,stenger1999b,bloch2000a,pitaevskii2003a,bloch2008a}. 

Position measurements erase the spatial phase coherence present before the measurement, at least on spatial scales comparable or greater than the measurement resolution. The dynamics of the associated decoherence process is an important question that also emerged during the development of quantum mechanics, \textit{e.g.} in the celebrated Heisenberg microscope thought experiment\,\cite{schiffQM}, and has been revisited many times (see, \textit{e.g.}, Refs.\,\cite{ozawa2003a,lund2010a,rozema2012a,busch2013a} for recent discussions and experiments). A simple but experimentally relevant example is a single atom illuminated by near-resonant light of wavelength $\lambda_0$. Here, light scattering from the incident mode to an initially empty mode can be interpreted as a weak, continuous measurement of the position of the atom\,\cite{marte1993a,gagen1993a,cirac1994a,pfau1994a,holland1996a}. Indeed, collecting the scattered photons with a suitable imaging system allows one (at least in principle) to infer the location of the source, \textit{i.e.} the atom. As more photons are scattered as a single realization of the experiment unravels, the position distribution of the atom (initially completely uncertain if the atom starts in a momentum eigenstate) progressively narrows down to a particular random location. Upon averaging over many realizations of the experiment, one retrieves a uniform average position distribution. In terms of the SPDM, the pinning of the particle position erases the spatial coherence of the initial state, leading to an exponential damping of $\rho^{(1)}(\bm{s})$ as long as $\vert \bm{s}\vert \gg \lambda_0$. In this elementary example, the damping rate is simply given by the rate of spontaneous emission $\gsp$ (or equivalently, the photon scattering rate). 

Instead of interferometry, the spatial coherence of a quantum system can also be characterized by measuring its momentum distribution $n(\bm{p})$\,\cite{cct2001a}, since $n(\bm{p})$ is the Fourier transform of the SPDM and therefore carries the same information\,\,\cite{stenger1999b,cct2001a,pitaevskii2003a}. A single photon scattering event corresponds to the absorption of an incident photon with wavevector $\bmkL$ and the subsequent emission of another photon with wavevector $k_ 0 \bm{u}$, where $k_0 \simeq \kL$ and where the unit vector $\bm{u}$ is random. Due to momentum conservation of the combined atom-light system, the atomic momentum changes from $\bm{p}_0$ to $\bm{p}_0 + \hbar(\bmkL - k_0 \bm{u})$. When many photons are scattered, the atomic state undergoes a random walk in momentum space, a phenomenon often called momentum diffusion. The characteristic width of the momentum distribution increases diffusively as $ \Delta k \propto \sqrt{t}$ for long times. Momentum diffusion plays an essential role in the theory of laser cooling, where it limits the achievable temperatures\,\cite{wineland1979a,gordon1980a}. In this context, the momentum diffusion of a single particle has been thoroughly studied.

In a recent work\,\cite{bouganne2020a}, we have investigated the decay of spatial coherence of bosonic gases in optical lattices when illuminated by a near-resonant excitation laser. Our main observation is a power-law decay of the coherence of strongly interacting systems at long times, rather than the exponential decay expected for an ensemble of non-interacting atoms undergoing momentum diffusion. Our observations are in quantitative agreement with the theoretical predictions of Poletti \textit{et al.}\,\cite{poletti2012a,poletti2013a} based on a minimal model describing light scattering by a quantum gas, hereafter denoted by ``dissipative Bose-Hubbard model'' (see Section\,\ref{sec:DBHM} for precise definitions). This model explains the power-law decay of coherence as the signature of a non-trivial dissipative many-body dynamics, where the interplay between fluorescence (destroying spatial coherence), and coherent tunnelling between neighboring sites (restoring spatial coherence at short distances) is drastically modified by interactions. In the present article, we extend our previous work in two respects. First, we focus on one-dimensional systems, as opposed to the two-dimensional systems studied in  \cite{bouganne2020a}. Second, we measure the spatial coherence between nearest-neighbor (n-n)  sites of the optical lattice (as in \cite{bouganne2020a}) but also between next-nearest-neighbor (n-n-n) sites. 

The article is organized as follows. In Section\,\ref{sec:setup}, we describe the experimental setup and protocols and review the essential features of a theoretical description of light scattering by atoms trapped in optical lattices, ignoring many-body effects for the time being. In Section\,\ref{sec:exp}, we present our main experimental results concerning the measurement of n-n and n-n-n correlators in one-dimensional lattice gases. In Section\,\ref{sec:DBHM}, we review the solution of the dissipative Bose-Hubbard model proposed by Poletti \textit{et al.}. In Section\,\ref{sec:C1}, we apply these results to the computation of n-n and n-n-n correlators measured experimentally. Finally, we conclude in Section\,\ref{sec:conclusion}.

\section{Experimental system}
\label{sec:setup}

\begin{figure}[!ht]
	\centering
	\includegraphics[width=\columnwidth]{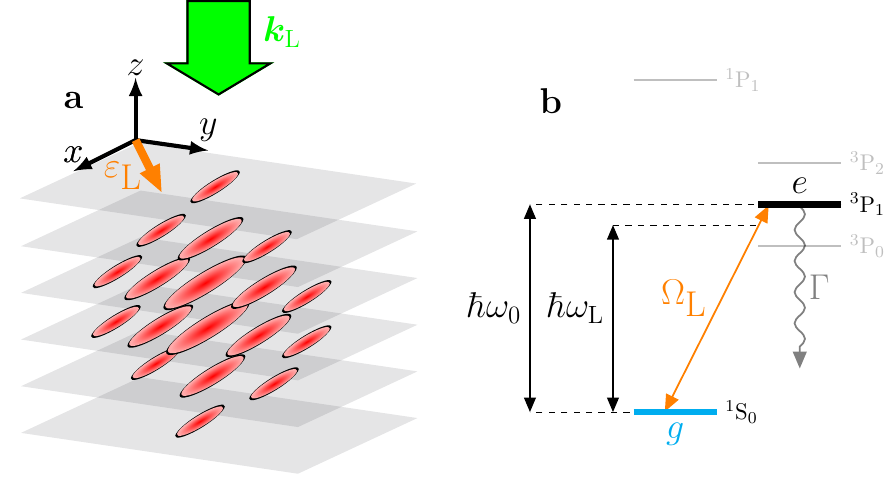}
	\caption{\textbf{a:} Sketch of the experimental geometry. Here $\bm{k}_{\mathrm{L}}$ and $\bm{\varepsilon}_{\mathrm{L}}$ denote the wave and polarization vectors characterizing the mode of the excitation laser.  \textbf{b:} Simplified atomic level structure of $^{174}$Yb highlighting the two relevant energy levels $g$ and $e$. The level splitting is $\hbar \omega_0$, the natural linewidth of the excited state is $\Gamma$, and the Rabi frequency controlling the strength of the interaction between the atom and the excitation laser is $\OmegaL$. The laser detuning from the resonance is defined as $\dL=\wL-\omega_0$.  }
	\label{fig1}
\end{figure}

\subsection{Experimental setup}

We start with a brief description of the experimental apparatus and methods (more details are given in \cite{dareau2015a,bouganne2017a}). We prepare a quantum gas of bosonic $^{174}$Yb atoms in a three-dimensional cubic optical lattice. We start from a nearly pure (more than 80\,\% condensed fraction) Bose-Einstein condensate (BEC) prepared in a crossed optical dipole trap\,\cite{dareau2015a}. We then ramp up the depth of the optical lattice potential along the vertical direction to a fixed value $V_z=27\,\ER$. Here $\ER=h^2/(2 \mYb \lambdaLat^2) \approx h \times 1.8\,$kHz is the recoil energy at the lattice wavelength $\lambdaLat\simeq 759.8\,$nm and $d=\lambdaLat/2$ is the lattice spacing. Simultaneously, we ramp down the depth of the crossed dipole trap to a negligible value and then switch it off once the ramp is completed. This sequence creates a stack of two-dimensional bulk Bose gases with negligible tunnelling between adjacent planes. In a second step, we increase the depths $V_{x,y}$ of the two horizontal lattices to their final values. In this work, we prepare arrays of independent one-dimensional (1d) lattice gases, as illustrated in Fig.\,\ref{fig1}\textbf{a}. We thus choose $V_x \ll V_y\approx V_z$ to enable tunnelling only in the $x$ direction. 

The density profile of the initial condensate is not uniform\,\cite{pitaevskii2003a} and the Gaussian envelopes of the lattice lasers result in an auxiliary harmonic confinement in the $x-y$ plane\,\cite{bloch2008a}. Both effects lead to an inhomogeneous distribution of atoms across all individual 1d gases\,\cite{paredes2004a,stoeferle2004a}. Varying the total atom number changes the peak density and the sizes of the initial Thomas-Fermi condensate, thereby allowing us to tune the peak density of our system and prepare a wide range of density profiles\,\cite{bouganne2017a}. Experiments presented in this article are performed with gases containing $N \approx 5 -6 \cdot 10^4$ atoms, where unit-, double- and triple-occupations coexist in the regime of deep lattices\,\cite{bouganne2017a}.

\subsection{Controlled spontaneous emission}
\label{sec:controlse}
\begin{figure*}[!!!!ht]
	\centering
	\includegraphics[width=\textwidth]{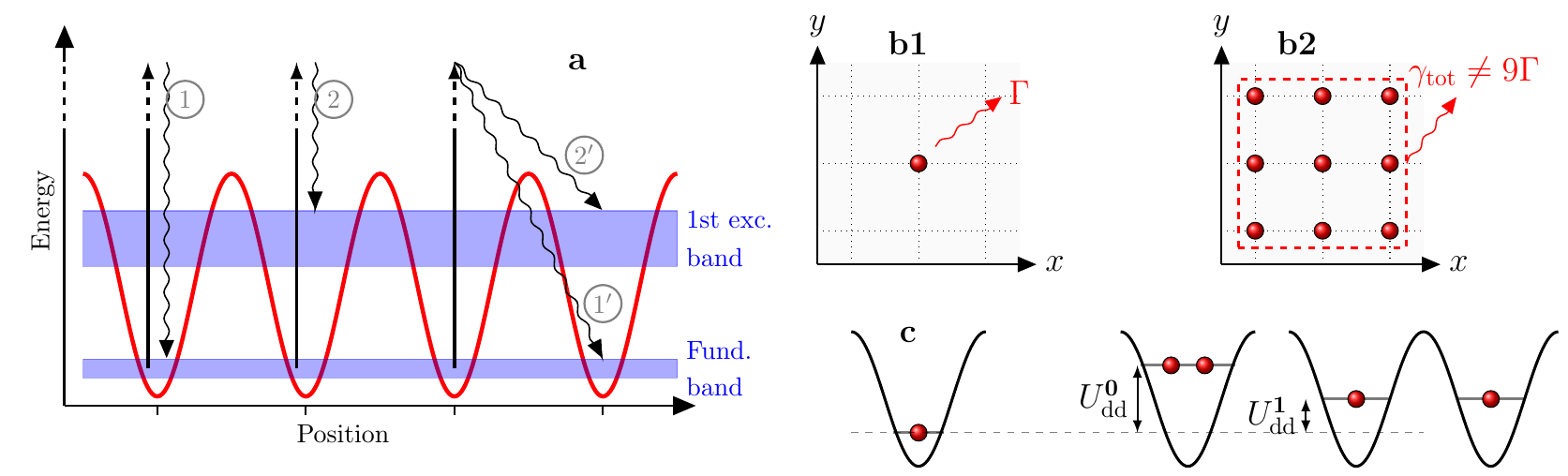}

	\caption{ {\textbf{a}:} Sketch of several possible transition processes after an absorption-spontaneous emission cycle. {Elastic scattering where atoms end up in the same band and site is denoted by (1). Same-site, band-changing processes such as the one labelled (2) are less frequent but not necessarily negligible for typical lattice depths ranging from a few $\ER$ to a few tens of $\ER$. In contrast, processes labelled (1'),(2') that involve tunnelling to a different site are exponentially suppressed due to the reduced overlap between adjacent Wannier states.} {\textbf{b1-2}:} Illustration of a single radiating atom (\textbf{b1}) and of collective radiation of an ensemble of nine atoms forming a cartoon Mott insulator state with one atom per site (\textbf{b2}). {\textbf{c}:} Illustration of energy shifts due to on-site ($U_{\mathrm{dd}}^{\bm{0}}$)
and nearest-neighbor ($U_{\mathrm{dd}}^{\bm{1}}$) dipole-dipole interactions. }
	\label{fig2}
\end{figure*}

In the absence of near-resonant laser light, the coherence of the gas is long-lived. Spontaneous emission induced by the far-detuned lattice lasers\,\cite{grimm2000a} corresponds to heating times of several seconds\,\cite{gerbier2010b,pichler2010a}, making spontaneous emission hardly distinguishable from other heating processes, \textit{e.g.} due to intensity fluctuations of the lattice lasers. To study the impact of spontaneous emission in a controlled way, we use an additional excitation laser tuned near the $^1\mathrm{S}_0 - ^3$$\mathrm{P}_1$ resonance at $\lambda_{\mathrm{0}} \approx 556\,$nm (see Fig.\,\ref{fig1}\textbf{b}). The actual detuning of the laser from the atomic resonance, $\dL=\wL-\wat \approx (2\pi)\times 2\,$MHz  with $\wat =c k_0$ the resonance frequency and $k_0=2\pi/\lambda_0$, is one order of magnitude larger than the spontaneous linewidth $\Gamma \approx (2\pi)\times180\,$kHz. We thus work in the regime of weak optical saturation, where the spontaneous emission rate is given by\,\cite{cct_atomphoton}
\begin{align}\label{eq:gsp}
\gsp \simeq \frac{\Gamma}{2}\frac{\Omega_{\mathrm{L}}^2}{2\delta_{\mathrm{L}}^2 + \frac{\Gamma^2}{2}} \ll \Gamma.
\end{align}
We choose the dipolar electric coupling strength $\OmegaL$ (controlled by the excitation laser intensity) such that $\gsp\approx 520\,\mathrm{s}^{-1}$ for the experiments presented in this article.

We now discuss how the motional quantum state of a single atom (or, equivalently, of a non-interacting gas of atoms) trapped in an optical lattice is modified by spontaneous emission when the gas is exposed to near-resonant laser light. The periodic optical lattice potential $V_{\mathrm{lat}}$ is created by a set of  far-off-resonance laser beams distinct from the excitation laser\,\cite{zwerger2003a,bloch2008a}. Because of the periodic potential, the motional levels cluster in allowed energy bands labeled by a band index $\nu$, with Bloch energy eigenstates  within each band also labeled by their quasi-momentum $\bm{q}$ restricted to the first Brillouin zone (BZ). In typical quantum gases experiments, only the fundamental energy band is occupied before exposing the atoms to near-resonant light. Once dissipation is enabled, light scattering leads to either intra-band or inter-band transitions depending on whether or not the band index $\nu$ changes. After a fluorescence cycle, the quasi-momentum is conserved  up to an arbitrary reciprocal lattice vector  $\bm{Q}$, \textit{i.e.} $\bm{q} \to  \bm{q} +\bmkL -k_0 \bm{u} +\bm{Q}$. With this minor modification, the phenomenon of momentum diffusion as described in the Introduction generalizes almost directly to the lattice case\,\cite{pichler2010a}.

For deep lattices and in the presence of interactions, it is often useful to use spatially localized Wannier states instead of (spatially delocalized) Bloch energy eigenstates\,\cite{bloch2008a}. The various types of transitions between Wannier states are illustrated in Fig.\,\ref{fig2}\textbf{a}. By far the dominant process is elastic scattering denoted by (1), where atoms end up in the same band and site, whereas same-site band-changing processes such as the one labelled (2) are much less frequent. Let us consider for concreteness interband transitions from the fundamental band to one of the first excited bands labeled by $\alpha=x,y,z$ in a cubic lattice. The transition rates scale as $\gsp \eta_\alpha^2$, where the so-called Lamb-Dicke parameter $\eta_\alpha \approx \kL \sigma_\alpha $ (with $\sigma_\alpha$ the harmonic oscillator width approximating the exact Wannier functions\,\cite{bloch2008a}) decreases slowly with increasing lattice depth. In contrast, processes involving tunnelling to a different site -- such as the ones marked (1'),(2') in \ref{fig2}\textbf{a} -- are exponentially suppressed due to the reduced overlap between adjacent Wannier states, and thus negligible in practice. A single-band description neglecting all bands but the fundamental one corresponds to the idealized limit where $\eta_\alpha \to 0$ (which occurs for infinite lattice depths). Lamb-Dicke parameters in our experiments are in the range $\eta_{y/z} \sim 0.1$ and $\eta_x \sim 0.2-0.4$. As a result, one expects that interband transitions take place on a time scale $\sim 10\, \gsp^{-1}$, as confirmed by experiments\,\cite{bouganne2020a}.

\subsection{Analysis of momentum profiles}

\begin{figure*}[hhhtttt!!!!!]
\centering
\includegraphics[width=\textwidth]{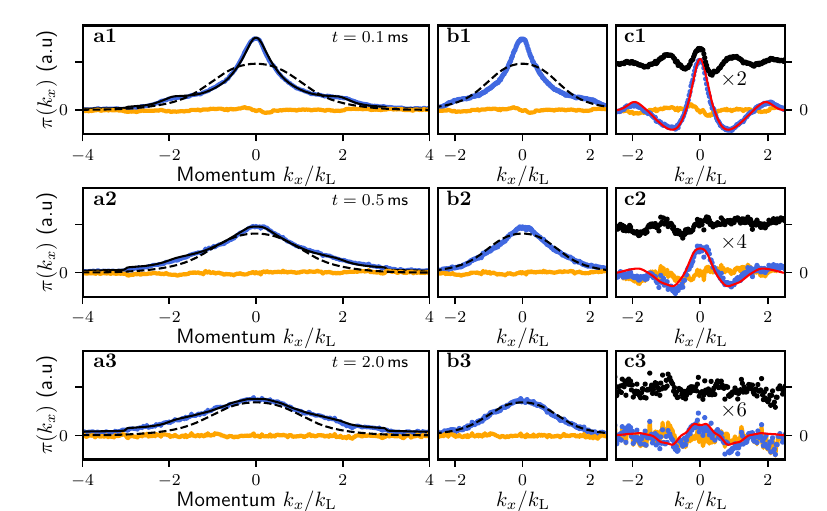}
\caption{{Extraction of nearest- and next-nearest-neighbors correlators $C_1$ and $C_2$ from the measured momentum distributions (here for a lattice depth $V_x =7.3\,\ER$ and $t = 0.1,0.5,2\,$ms). The first Brillouin zone corresponds to $[-\kL,+\kL[$. The blue solid lines in the left panels \textbf{a1-3} show the experimental data for the normalized momentum distribution $\pi(k_x)$, the black solid lines the best fit using Eq.\,(\ref{eq:nkexp}), and the orange lines the fit residuals. In the middle panels \textbf{b1-3}, the blue dots show the contribution of the fundamental band $\pi_0=\pi(k_x)-f_{\mathrm{exc}}$ obtained by subtracting from the measured $\pi$ the content $f_{\mathrm{exc}}$ of the fit function describing the excited bands [Eq.\,(\ref{eq:nkexp})]. The dashed lines in the left and middle panels show the ``incoherent'' part of the fundamental band contribution, \textit{i.e.} the calculated Wannier envelope weighted by the band population $p_0 \mathcal{W}_0(k_x)$. In the right panels \textbf{c1-3}, the blue dots show $\pi_0-p_0 \mathcal{W}_0(k_x)$, the black dots (offset vertically for clarity) show $\pi_0-p_0 \mathcal{W}_0(k_x)- 2 C_{1}\cos( k_x d )$, and the red solid line is the fitted curve $2 \mathcal{W}_0(k_x) \big( C_{1}\cos( k_x d ) +C_{2}\cos(2k_x d )\big)$. The left and middle panels use the same (arbitrary) vertical scales. The vertical scales of the rightmost panels are enlarged by the factors indicated in the plots to make the details more visible.}}
\label{fig3}
\end{figure*}

{All measurements in this article are done using absorption imaging of the atomic gas. We record the images after releasing the gas from the optical lattice trap and letting it expand for 20\,ms. Such absorption images allow one to infer the atomic momentum distribution integrated along the probe line  of sight, $
\int n(\bm{k},t) dk_z$. As in our previous work\,\cite{bouganne2020a}, we use a model function to fit the experimental momentum distributions. We first recall, for completeness and to motivate our fit model, a few key notions concerning the momentum distribution of a lattice gas, before discussing our fitting model in details.}

 \paragraph{Definitions of single-particle correlators and quasi-momentum distribution:}
For a one-dimensional quantum gas occupying the fundamental Bloch band $n=0$, the discrete single-particle correlator\,\cite{greiner2001a,pedri2001a,zwerger2003a}
\begin{align}\label{eq:Cij}
C_{s}= \frac{1}{N}\sum_{i} \langle \hat{a}_{i+s}^\dagger \hat{a}_i \rangle
\end{align}
plays the same role as the SPDM $\rho^{(1)}$ for a gas in the continuum. Here $\hat{a}_i$ is the  operator annihilating an atom at site $i$ with position $x_i=i\cdot d$, $s$ is the relative distance in units of the lattice spacing $d$, and $N$ is the total atom number. In the case of a uniform system with $N_s$ sites and filling factor $\bar{n}=N/N_s$, the general definition (\ref{eq:Cij}) simply amounts to a normalization such that $C_{0} =1$, \textit{i.e.} $C_{s}= \langle \hat{a}_{i+s}^\dagger \hat{a}_i \rangle /  \bar{n}$. The definition (\ref{eq:Cij}) is more general and applies in particular to systems with a non-uniform density, as realized in experiments. The one-dimensional momentum distribution\,\cite{pedri2001a,zwerger2003a,bloch2008a}
\begin{align}\label{eq:nk0}
n_0(k_x)=  \mathcal{W}_0(k_x) \left( 1 +  \sum_{s \in \mathbb{N}^\ast} 2C_{s} \cos( s k_x d) \right)
\end{align}
is proportional to the product of a ``structure factor'', the discrete Fourier transform of $C_{s}$, by an envelope function $\mathcal{W}_0(k_x)$. The latter is the square of the Fourier transform of the real-space Wannier function in the fundamental band and reflects the momentum distribution of a single atom confined at one lattice site. The structure factor can also be interpreted as a normalized quasi-momentum distribution. When taking excited bands into account, Eq.\,(\ref{eq:nk0}) is generalized as $n(k_x)=n_0(k_x)+n_{\mathrm{exc}}(k_x)$, with $n_{\mathrm{exc}}(k_x)= \sum_{\mathrm{exc.\,\,bands}\,\, n} \mathcal{W}_n(k_x) \left( 1 +  \cdots \right)$
the contribution of the excited bands.

\paragraph{Model function:}

We integrate the atomic distributions $\int n(\bm{k},t) dk_z$ inferred from the absorption images over the irrelevant direction $k_y$, and normalize them to the total atom number,
\begin{align}\label{eq:piexp}
\pi(k_x,t) = \frac{1}{N_{\mathrm{tot}}(t)} \int n(\bm{k},t)dk_y dk_z.
\end{align}
Experiments presented in this article are performed with one-dimensional systems characterized by short-ranged spatial coherence even in the equilibrium state, due to the central role played by fluctuations in one-dimensional systems\,\cite{cazalilla2011a}. As a result, the Fourier expansion in Eq.\,(\ref{eq:nk0}) can be truncated to the first few terms. We fit the distribution $\pi$ with $f(k_x,t) = f_0(k_x,t)+f_{\mathrm{exc}}(k_x,t)$, where the contribution of atoms in the fundamental band and excited bands are respectively
\begin{align}\label{eq:nkexp}
f_0(k_x,t)&=\mathcal{W}_0(k_x) \Big( p_0 + 2 \sum_{s=1}^{3} C_{s}(t)\cos( s k_x d ) \Big),\\ 
f_{\mathrm{exc}}(k_x,t) &=\sum_{n=1}^{n_{\mathrm{max}}} \mathcal{W}_n(k_x) p_n,
\end{align}
with $p_n$ the fractional  population of band $n$. The fundamental band correlators $C_s$  are the same as in Eq.\,(\ref{eq:Cij}). The definitions are such that $\int \pi(k_x,t) dk_x =1$, $\sum_n p_n=1$. In the excited bands contribution, we neglect all correlators and use a cutoff $n_{\mathrm{max}}=3$. We calculate the Wannier envelope functions $\mathcal{W}_n(k_x)$ at the lattice depths of interest, and use them as input to the fitting procedure. 
The free parameters in the fit are thus the fundamental band correlators $\{C_s\}_{s=1,\cdots,3}$ and the band populations $\{p_n\}_{s=0,\cdots,3}$.

We show in Fig.\,\ref{fig3}\textbf{a1-3} three illustrative momentum profiles along with the best fit and the fit residuals. Comparing the measured distribution with the lowest band Wannier enveloppe (shown as dashed lines) at various times, the broadening of the overall distribution reflects the gradual population of the excited bands due to the excitation laser, as discussed in Section\,\ref{sec:controlse}. The fit residuals are featureless and the amplitude of the residual deviations is comparable to the detection noise. We conclude that the fit function is able to capture the relevant features of the measured momentum distribution, and in particular to substract efficiently the contribution of the higher bands. To appreciate the importance of the various terms, we show the contribution $\pi - f_{\mathrm{exc}}$ of the fundamental band in Fig.\,\ref{fig3}\textbf{b1-3}. This quantity can be compared to the momentum distribution of a band with the same overall population but no coherence, shown as dashed lines. Fig.\,\ref{fig3}\textbf{c1-3} shows the extracted interference pattern from atoms in the fundamental band. When $C_1 \neq 0$, the probability to find atoms near $k_x =0,\pm 2 \kL$ increases due to constructive interference, while the probability near $k_x = \kL$  decreases due to  destructive interference. Note that the Wannier enveloppe suppresses the visibility when $k_x \neq 0 $. Subtracting the n-n contribution from the interference patter, we obtain the black dotted lines where the additional modulation with period $\kL$ comes from n-n-n coherence.

\section{Observation of anomalous decay of coherence in one dimension}
\label{sec:exp}

\subsection{Observation of algebraic decay at long times}

\begin{figure*}[hhhtttt!!!!!]
\centering
\includegraphics[width=\textwidth]{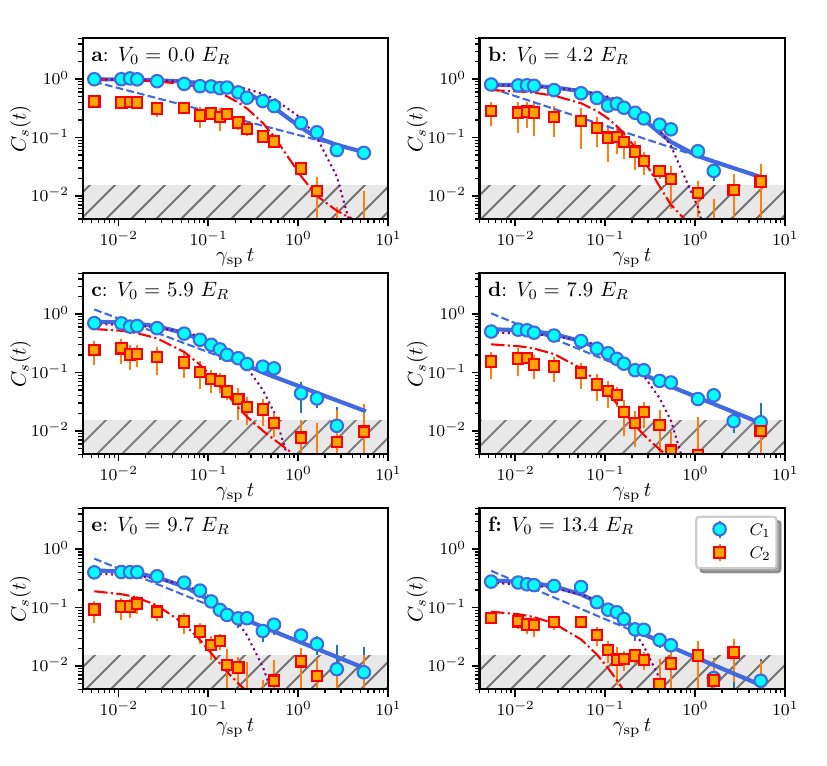}
\caption{Nearest- (circles) and next-nearest (squares) neighbor correlators $C_1$ and $C_2$ as a function of time for several lattice depths indicated in the individual panels \textbf{a}-\textbf{f}. The blue dotted and solid lines show respectively an exponential fit to $C_1(t)$ and an improved fit using a function [Eq.\,(\ref{eq:C1fit})] that first decays exponentially before crossing over to an asymptotic power-law decay. The asymptotic power law itself is shown by the dashed line. The theory based on the dissipative Bose-Hubbard model (Section\,\ref{sec:DBHM}) predicts the asymptotic law $C_2 \approx C_1^2$. We test this prediction by taking the square of the improved fit to $C_1$ without any additional change (red dot-dashed lines). We observe a reasonable agreement with the experimentally measured $C_2$ (red squares) at sufficiently long times. The hashed gray areas  show the regions $C_s < 0.02$ where detection noise becomes dominant. These regions are excluded from the fits. The experimental error bars are statistical. We record typically  five independent realizations per data point $(V_x,t)$. We fit each realization independently and report the mean and standard deviation of the fitted $C_{1/2}$. }
\label{fig4}
\end{figure*}

Fig.\,\ref{fig4} shows the evolution of the nearest- (circles) and next-nearest (squares) neighbors correlators $C_1$ and $C_2$ as a function of the dissipation time for several values of the lattice depth $V_x$, that decay as expected. The hashed gray areas in Fig.\,\ref{fig4} show the detection noise, \textit{i.e.} the imaging noise in absorption images, mostly due to the optical shot noise of the probe laser. Values of $C_s$ below a threshold of roughly $10^{-2}$ are dominated by noise and therefore not indicative of atomic coherence. We use a conservative threshold value $C_s \geq 0.02$, excluding data points with lower coherence from the fits. The dashed gray areas in Fig.\,\ref{fig4} show the excluded regions. Even at short times, the fitted third-nearest neighbor coherence $C_3$ are typically barely distinguishable from the detection noise. We have thus omitted $C_3$ from the figures to improve their readability. Furthermore, the detection noise restricts the observation times to $\gsp t \lesssim 5$ to maintain a sufficient signal-to-noise ratio for $C_{1/2}$ for all lattice depths. 

With suitable approximations, the quantum optical theory of relaxation can be used to predict the evolution of the momentum distribution\,\cite{pichler2010a,poletti2012a,poletti2013a,yanay2014a}. The simplest model neglects all correlations from interatomic interactions or collective effects in the radiation of the atomic ensemble: Each atom interacts with the excitation laser and the vacuum field independently from the presence of others. Assuming that only the fundamental Bloch band is relevant, this model predicts that the initial coherences decay exponentially (see\,\cite{pichler2010a,yanay2014a} and Appendix\,\ref{app:theoryRelax}),
\begin{align}\label{eq:Cs_independent}
C_s(t) \simeq C_s(0) \exp (- \gsp t) .
\end{align}
This prediction is in poor agreement with the experimental results. Although the initial decay of the measured coherences is well described by an exponential function (dotted lines in Fig.\,\ref{fig4}), the fitted decay rates are substantially larger than the value $\gsp$ suggested by Eq.\,(\ref{eq:Cs_independent}), see Fig.\,\ref{fig5}c. Furthermore, the experimental curves eventually depart from the exponential behavior and settle at long times to a regime where the decay becomes slower than predicted by Eq.\,(\ref{eq:Cs_independent}). In that regime, the experimental data tend to align as a straight line in a double-logarithmic plot, indicating that the long-term decay follows an algebraic law $\propto t^{-\alpha}$ instead of an exponential one.

\subsection{Atom losses}

In addition to the decay of coherence and to interband transfer, we also observe an overall decay of the total atom number due to the excitation laser. {We attribute these losses to light-induced two-body collisions converting a fraction of the internal energy into kinetic energy and leading to atom losses\footnote{A contribution from nearby photoassociation resonances is also possible.}. This observation is qualitatively and quantitatively in line with our previous work on two-dimensional systems\,\cite{bouganne2020a}, where it was analyzed in details. The loss time constant is roughly one order of magnitude slower than the decoherence dynamics.} The main effect of inelastic losses --the reduction of atom number-- is normalized away by the definition (\ref{eq:Cij}) of the single-particle correlators. However, there remains a more subtle effect on the coherence that will be discussed later in Section\,\ref{sec:C1}.

\subsection{Determination of the decay exponent}

\begin{figure*}[hhhtttt!!!!!]
\centering
\includegraphics[width=\textwidth]{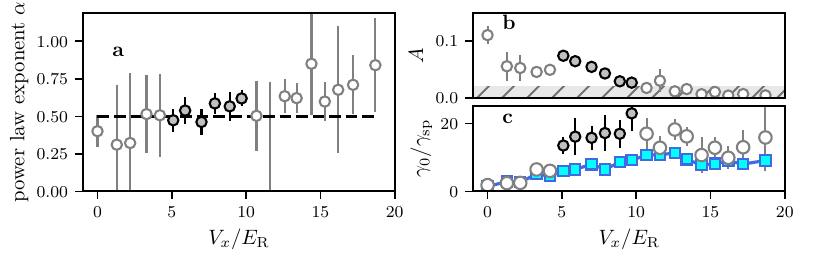}
\caption{Results of fitting Eq.\,(\ref{eq:C1fit}) to the experimental data: \textbf{a:} Power law exponent $\alpha$; The dashed line in \textbf{a} shows the decay exponent $\alpha=1/2$ expected from theory [section\,\ref{sec:DBHM}]; \textbf{b:} power law amplitude $A$; \textbf{c:} initial decay rate $\gamma_0$ plotted as a function of lattice depth. The filled circles correspond to measurements made in the interval $5\,\ER \leq V_x\leq 10\,\ER$ where the exponent can be reliably extracted, while open circles show measurements outside this interval. The average exponent for $5\,\ER \leq V_x\leq 10\,\ER$ is $\alpha \approx 0.54(6) $. The blue squares in \textbf{c} are the decay rate of a single exponentially decaying function $C_1(0)\ee^{-\gamma_0 t}$ fitted to the data. All error bars are one standard deviation confidence intervals estimated from the fit.}
\label{fig5}
\end{figure*}

We now analyze in more details the behavior of the single-particle coherences $C_1$ and $C_2$ shown in Fig.\,\ref{fig4}.  The theory presented in Section\,\ref{sec:DBHM} explains the origin of the asymptotic power-law tail (shown as dashed line in Fig.\,\ref{fig4}).  {However, the same theory only predicts the asymptotic behavior and cannot describe the transient behavior at short times $\gsp t \lesssim 1$, where the initial coherence are quickly suppressed before the power-law regime emerges\,\cite{poletti2012a}.} Experimentally, we are able to explore a relatively narrow time window $0 \leq \gsp t \lesssim 5$ due to the parasistic loss processes mentionned above. {As a result,} modelling this transient behavior is important to extract a reliable value for the power-law exponent. 

The solid curve in Fig.\,\ref{fig4} shows a fit to the experimentally measured $C_1(t)$ using the function
\begin{align}\label{eq:C1fit}
\mathcal{C}_1(t)=& C_1(0) \exp(-\gamma_0 t) +  A \frac{1 - \exp(-\gamma_0 t) }{(\gsp t)^\alpha}. 
\end{align}
Eq.\,(\ref{eq:C1fit}) should be viewed as a heuristic function describing a smooth crossover from an initially exponential decay to a long-time algebraic decay with exponent $\alpha$. For simplicity, we have chosen to use the same exponential function to describe the initial decay and the build up of power-law behavior. We use the initial decay rate $\gamma_0$, the asymptotic amplitude $A$ and the decay exponent $\alpha$ as free fitting parameters.

The results of the fit are presented in Fig.\,\ref{fig5}. The main result concerns the exponent $\alpha$ of the power law, plotted in Fig.\,\ref{fig5}\textbf{a}. The limitations on detection sensitivity and observation times restrict the range of lattice depths where a power-law tail can be reliably extracted to intermediate values, roughly $5\,\ER \leq V_x \leq 10\,\ER$. For clarity of presentation, we have marked with open symbols in Fig.\,\ref{fig5} measurements outside of this ``interval of reliability''. For lattice depths below $ \sim 5\, \ER$, the observed decay is essentially described by the exponential part, although a slower tail at long times seems to develop in all cases (even without a lattice). For such low lattice depths, the fitted initial decay rate $\gamma_0$ (circles in  Fig.\,\ref{fig5}\textbf{c}) is almost identical to the decay rate of a single exponentially decaying curve fitted to the data (blue squares). For lattice depths above $\sim 12\,\ER$, the lattice dynamics becomes slower than the spontaneous emission time, $\hbar/J \geq \gsp^{-1}$. As a result, when the power law tail eventually appears, its amplitude is close or below the noise floor (see Fig.\,\ref{fig5}\textbf{b}) and extracting reliably  the exponent becomes difficult. This behavior is reflected by increased uncertainties on the fitted power law exponent outside the interval $5\,\ER \leq V_x \leq 10\,\ER$. Inside the ``interval of reliability'', we find an average value $\alpha \approx 0.54(6) $ consistent with $\alpha = 0.5$.

Another observation in Fig.\,\ref{fig4} concerns the next-nearest-neighbor correlator $C_2$. We find that $C_2$ is well described by $\mathcal{C}_1^2$ for $t \gg \gamma_0^{-1}$, where $\mathcal{C}_1$, given in Eq.\,(\ref{eq:C1fit}), is fitted to the nearest neighbor coherence without additional adjustment parameter. This behavior is consistent with an exponential decay of the correlator $C_s \sim e^{-s}$ for $s \neq 0$. In the next Section\,\ref{sec:DBHM}, we analyze a minimal model that explains the two effects that we observe experimentally, namely the emergence of a power law behavior $C_1 \propto 1/ t^{1/2}$ and the relation $C_2 \approx C_1^2$.

\section{The dissipative Bose-Hubbard model}
\label{sec:DBHM}

For the sake of simplicity in the following discussion, we  restrict ourselves to a quantum gas in the fundamental band $\nu_0$, ignoring the possibility of interband transfer. Extending the theory given below to take multiple bands into account is straightforward in principle, but leads to a substantial increase of the complexity. Ref.\,\cite{pichler2010a} establishes the minimal model describing the dynamics of a gas of bosonic atoms in an optical lattice illuminated by near-resonant light. This model can be described by a Lindblad master equation of the form 
\begin{align}\label{eq:lindblad}
\frac{d}{dt}\hat{\rho}  & = \frac{1}{i\hbar}\big[ \hat{H} , \hat{\rho} \big]  + \mathcal{K}\big[\hat{\rho}\big].
\end{align}
Here $\hat{H}$ denotes the Bose-Hubbard (BH) Hamiltonian, 
\begin{align}\label{eq:H}
\hat{H}  & =-J \sum_{\langle i,j \rangle} \hat{a}_{i}^\dagger \hat{a}_{j}
+
\frac{U}{2} \sum_{i} \hat{n}_{i}( \hat{n}_{i}-1),
\end{align}
with $J$ and $U$ the nearest-neighbor tunnelling and interaction energies. The superoperator $\mathcal{K}$ (hereafter called ``dissipator'') reads
\begin{align}
\label{eq:K}
\mathcal{K}\big[\hat{\rho}\big] & = \gsp \sum_{i }\hat{n}_{i} \hat{\rho} \hat{n}_{i} - \frac{1}{2} \Big\{ \hat{n}_{i}^2, \hat{\rho} \Big\}.
\end{align}
We refer to the model defined by Eq.\,(\ref{eq:lindblad}), $\hat{H}$ and $\mathcal{K}$ as  ``dissipative Bose-Hubbard model'' in the following. 
In Appendix\,\ref{app:theoryRelax}, we discuss $N-$particle corrections, either collective radiative effects where the spontaneous emission rate of $n$ atoms differs from $n \gsp$ (illustrated in Fig.\,\ref{fig2}\textbf{b}), or long-range interactions between the induced electric dipoles carried by each atom mediated by multiple photon scattering (illustrated in Fig.\,\ref{fig2}\textbf{c})\,\cite{pichler2010a}. We conclude that collective effects are negligible for the experimental parameters used in this work. 

The dissipative BH model has been analyzed by Poletti \textit{et al.}\,\cite{poletti2012a,poletti2013a}. After an initial stage for $\gsp t \leq 1$, where spatial coherences eventually present in the initial state are exponentially damped, they discovered an asymptotic regime where relaxation slows down dramatically. The theory of \cite{poletti2013a} is summarized in 
the Supplementary Material.
The key result is a classical master equation for the on-site number distribution $p_n$, \textit{i.e.} the probability to find $n$ bosons at any given site,
\begin{align}\label{eq:masterEq}
\frac{d p_n}{d \tau} &= \sum_{\sigma = \pm 1} W_{n+\sigma}\big[ \{ p_n \} \big] \Big( p_{n+\sigma} - p_n \Big),
\end{align}
with $\tau=t/T^\ast$ a dimensionless time. The dimensionless transition rates are given by $W_{n+1}\big[ \{ p_n \} \big] =\sum_m g_{\varepsilon,\overline{n}}(m,n+1)p_{m-1}$, $W_{n-1}\big[ \{ p_n \} \big] = \sum_m g_{\varepsilon,\overline{n}}(m+1,n)p_{m+1}$, with the function $g_{\varepsilon, \overline{n}  }(m,n)= m n/[(m-n+1)^2+(\overline{n} \varepsilon)^2]$. The dynamics is governed  by the dimensionless dissipation parameter
\begin{align}
\varepsilon = \frac{\hbar \gsp}{U \overline{n}},
\end{align}
and by an emergent time scale $T^\ast$ such that
\begin{align}
\frac{1}{T^\ast}& = 2z\gsp  \left(\frac{J}{U \overline{n}}\right)^2.
\end{align}
Here $z=2$ is the number of nearest-neigbors in one dimension and $\bar{n}$ the filling factor (mean number of atoms per lattice site).

\begin{figure*}[!ht]
	\centering
	\includegraphics[width=\textwidth]{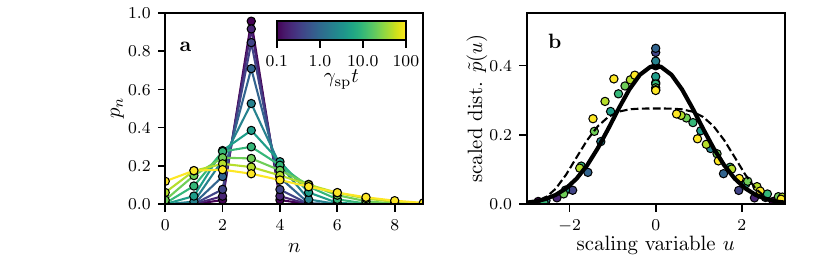}
	\caption{Numerical solution of the master equation (\ref{eq:masterEq}) for the on-site distribution $p_n$. {\textbf{a:}} bare distribution as a function of occupation number $n$. {\textbf{b:}} rescaled distribution $\tilde{p}(u)=p_n \times \bar{n} \tau^{\frac{1}{4}}$ as a function fo the scaling variable $u=(n-\bar{n})/(\bar{n} \tau^{\frac{1}{4}})$. The dashed line shows the scaling distribution $f_\infty$ and the solid line a normalized Gaussian function. We used the parameters $ U/J=20$, $\varepsilon=0.1$ and $\bar{n}=3$.}
	\label{fig6}
\end{figure*} 

A numerical solution of the master equation (\ref{eq:masterEq}) is shown in Fig.\,\ref{fig6}\textbf{a}. The initial state is a Fock state with $\overline{n}=3$. The master equation conserves the norm $\sum_n p_n = 1$ and the average population $\sum_n n p_n = \overline{n}$. The distribution function broadens with time as expected, but in a very peculiar way. Fig.\,\ref{fig6}\textbf{b} illustrates that the distribution function obeys approximately a scaling dynamics. The distribution functions at widely different times overlap when plotted versus the scaling variable\,\cite{poletti2013a},
\begin{align} \label{eq:scalingvar}
u\left( n/\bar{n},\tau\right)= \frac{n/\bar{n}-1}{ \tau^{\frac{1}{4}}}.
\end{align}
The scaling property implies that the standard deviation of $n$ increases with time roughly as a power law $\propto \tau^{1/4}$, much more slowly than for a standard diffusive process $ \propto \tau^{1/2}$. 

To gain further insight on the origin of this scaling behavior, we follow again Poletti \textit{et al.}\,\cite{poletti2013a} and consider the limiting case of very large filling $\overline{n} \gg 1$. In that limit, the discrete variable $n$ and distribution $p_n$ can be replaced by their continuous counterparts, $x=n/\overline{n} \in [0,\infty[$ and $p(x,\tau)=p_n(\tau)/\delta x$, with $\delta x = 1/\overline{n}$. The classical master equation (\ref{eq:masterEq}) maps to a diffusion equation\,\cite{poletti2013a},
\begin{align}\label{eq:diffeq}
\frac{\partial p(x,\tau)}{\partial \tau} \approx \frac{\partial}{\partial x} \Big( D(x) \frac{\partial p(x,\tau)}{\partial x} \Big).
\end{align}
with a diffusion coefficient $D(x) = 1/ [ \varepsilon^2+(x-1)^2]$.

The diffusion equation admits two limiting cases of interest, a dissipation-dominated regime $\hbar \gsp \gg U$ (\textit{i.e.} $\varepsilon\gg 1$), with a diffusion coefficient $D(x) \approx 1/\varepsilon^2$, and an interaction-dominated regime $U \gg \hbar \gsp$ (\textit{i.e.} $\varepsilon \ll 1$), with a diffusion coefficient $D(x) \approx (x-1)^{-2}$. In both cases, the diffusion coefficient behaves as a power law $D(x) \sim x^{-\eta}$ with $\eta=0$ ($\varepsilon\gg 1$) or $2$ ($\varepsilon\ll 1$). For the initial condition  $p(x,0)=\delta(x)$, the diffusion equation\,(\ref{eq:diffeq}) with a power law diffusion coefficient admits a self-similar solution\,\cite{poletti2013a}
\begin{align} \label{eq:p_realunits}
p(x,\tau) & = \tau^{-\beta} f_{\infty}\left[ u\left( x,\tau\right) \right],
\end{align}
with scaling exponent $\beta=1/(\eta+2$). For an arbitrary initial condition $p_0(x)$, the time evolution can be obtained from $p(x,\tau)=\int dx_0 p_0(x_0) G(x-x_0,\tau)$ owing to the linearity  in $p(x)$ of the diffusion equation.

In the dissipation-dominated regime $\varepsilon \gg 1$, the scaling solution $f_{\infty}$ is a Gaussian function corresponding to a scaling exponent $\beta=1/2$. This corresponds to standard diffusive behavior with $\Delta x \propto \sqrt{t}/\varepsilon$. The diffusion coefficient in real units $D_{\varepsilon} \sim 1/(T^\ast\varepsilon^2) \sim J^2/(\hbar^2\gsp) \ll J/\hbar$ describes a diffusive dynamics much slower than the ballistic dynamics of atoms tunnelling in the lattice. Such a slowed down, dissipative dynamics can be interpreted as resulting from the inhibition of coherent tunnelling by the quantum Zeno effect\,\cite{patil2015a}.

The interaction-dominated regime $\varepsilon \ll 1$ is characterized by a different scaling function $f_{\infty}( u) = \ee^{-16 u^4}/\Gamma(1/4)$ ($\Gamma$ is the Euler gamma function) and exponent $\beta  =1/4$. Numerical simulations as in Fig.\,\ref{fig5} show that even for finite $\bar{n} \geq 1$, the distribution function obeys approximately the scaling relation $p(x,\tau) \approx \tau^{-\beta} f_{\mathrm{approx}}\left[ u\left( x,\tau\right) \right]$ with exponent $\beta  \approx 0.25$. This approximate scaling behavior for finite $\bar{n}$ explains the sub-diffusive dynamics observed in the numerical calculation. However, we find empirically that the approximate scaling function for finite $\bar{n}$ differs from $f_ \infty$. Numerically, we find that it is closer to a normalized Gaussian function $f_{\mathrm{approx}} ( u) = \ee^{-u^2/2}/\sqrt{2\pi}$ (see Fig.\,\ref{fig6}\textbf{b}).

\section{Single-particle correlation function}
\label{sec:C1}
%
To connect the theory and the experimental results, we now turn to the calculation of the single-particle correlators. We find 
that the nearest-neighbor correlator depends on the populations as (see the Supplementary Material for details of the calculations)
\begin{align}
\nonumber C_1(t)  & \simeq   \frac{J}{U \bar{n}} \sum_{m,n=0}^{+\infty}\frac{m(n+1)(n-m+1)}{(n-m+1)^2+(\varepsilon\overline{n})^2}\\
& \label{eq:C_1}
 \hspace{1.5cm}\times \Big[ p_{m} p_{n} - p_{m-1} p_{n+1}\Big].
\end{align}
We have also computed the next-nearest-neighbors correlator $C_2(t)$ in terms of the distribution function $p_{n}$ using the same method as for $C_1$. We do not reproduce here the rather lengthy expression [see Eq.\,(15)
in the Supplementary Material]. 

When the distribution $p$ assumes a scaling form $p(n,\tau) \approx \tau^{-\beta} f\left[ u \right]$, the correlators $C_{1,2}$ obey simple laws (see Supplementary Material),
\begin{align}
\label{eq:C1scaling}
C_1(t) & \simeq \frac{a_1}{\sqrt{2 z \gsp t}},\\
\label{eq:C2scaling}
C_2(t)&  \simeq \frac{a_2}{2 z\gsp t} = \frac{a_2}{a_1^2} C_1^2(t).
\end{align}
The numerical coefficients $a_1,a_2$ depend on the particular form of the distribution function $f$. For the approximate but experimentally relevant Gaussian scaling distribution, we find $a_1 = a_2 = 1$ and
\begin{align}\label{eq:C2overC1sqr_scaling}
C_2(t) \underset{\bar{n} \sim 1}{\simeq}  C_1^2(t).
\end{align}
Thus, the scaling approach fully explains the experimental observations of Section\,\ref{sec:exp}.

\begin{figure*}[!ht]
\centering
\includegraphics[width=\textwidth]{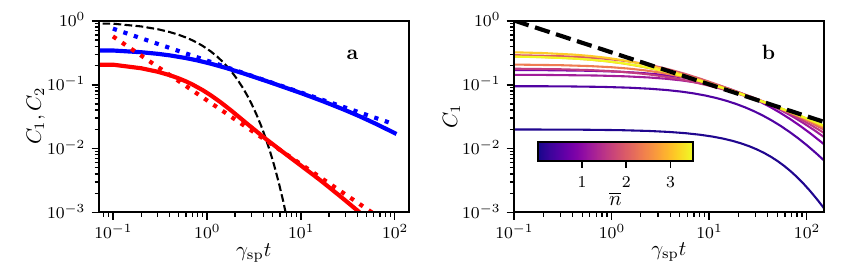}
\caption{ {\textbf{a}:} Nearest-and next-nearest-neighbor correlators $C_{1}$ (blue) and $C_2$ (red) evaluated for $U/J=20$ and $\epsilon=0.1$ with a Fock state with $\bar{n}=3$ as initial state. The numerical calculations (solid lines) are compared to the analytic scaling forms in Eqs.\,(\ref{eq:C1scaling},\ref{eq:C2scaling}) [dotted lines]. The dashed line shows the exponential decay curve expected for independent atoms. {\textbf{b}:} Calculation of $C_1$ for several filling factors $\bar{n} \in [0.1,3.5]$ and the same parameters as in \textbf{a}. The dashed line shows the asymptote predicted in the scaling limit $\bar{n}\to + \infty$, $\epsilon \to 0$. }
\label{fig7}
\end{figure*}

As for the distribution function, the numerical evaluations of $C_{1/2}$ shown in Fig.\,\ref{fig7} confirm the relevance of the scaling solution even for relatively small filling factors. Fig.\,\ref{fig7}\textbf{a} shows that the asymptotic laws in Eqs.\,(\ref{eq:C1scaling},\ref{eq:C2scaling}) describe well the algebraic tails of the correlators for $\bar{n} =3$. Fig.\,\ref{fig7}\textbf{b} concentrates on $C_1$, varying the filling factor in the interval $\bar{n} \in [0.1,3.5]$. When $\bar{n} \gtrsim 1$, the correlator $C_1$ approaches the scaling law (\ref{eq:C1scaling}) from $\gsp t \gtrsim 1$ up to a filling-dependent time $T_{\mathrm{sc}}$. Thus, one can define a ``scaling window'' $ t \in [\gsp^{-1}, T_{\mathrm{sc}}]$ where the correlator is well described by the scaling theory. Outside this window, the distribution has broadened sufficiently to ``feel the $n=0$ boundary'', the distribution deviates from the scaling form and the algebraic behavior disappears. 

To conclude this analysis, we briefly discuss how the overall picture of scaling dynamics is modified in presence of losses, as in the experimental system. We restrict ourselves to the ``adiabatic regime'' (approximately realized in the experiments), where the loss rate is much slower than the spontaneous emission rate. The scaling solution is still relevant, but with a slowly decaying $\bar{n}(t)$ due to losses. Fig.\,\ref{fig7}\textbf{b} then allows us to understand qualitatively that losses shorten the scaling window. For very small loss rates, the density barely changes in the entire scaling window, and the scaling picture of the evolution of coherence is therefore not significantly modified. As the loss rate is increased,  the system follows as time goes the set of curves shown in Fig.\,\ref{fig7}\textbf{b}, moving to the right due to spontaneous emission but also downwards towards lower filling factors because of losses. Eventually, the system exits the scaling regime not because of the initial finite duration of the scaling window, but because $\bar{n}(t)$ has dropped to a value $\sim 1$ where the scaling behavior disappears. Numerical simulations\,\cite{bouganne2020a} support the claim that our experiments are performed in this intermediate, adiabatic regime, where losses are slow enough to use an adiabatic description, but fast enough to impact the evolution of coherence.

\section{Conclusion}
\label{sec:conclusion}

In this article, we have given a detailed experimental and theoretical account of the loss of spatial coherence in one-dimensional bosonic quantum gases illuminated by near-resonant light. We analyze the momentum distribution to extract nearest- and next-nearest-neighbors correlators of the bosonic field. We find experimentally that the n-n correlator decays as a power law $\propto 1/\sqrt{t}$, and also that the n-n correlator controls the n-n-n correlator according to $C_2 \approx C_1^2$ at long times. Our findings are in contrast to the simple model where the atomic ensemble is treated as mutually independent scatterers, and where the decay would be exponential with a time scale given by the inverse of the spontaneous emission rate. The power law decay at long times is correctly predicted by the theory of \cite{poletti2012a}, which agrees quantitatively with our measurements in the sense that it predicts an exponent $\alpha = 1/2$ and the same relation $C_2 \approx C_1^2$ between n-n and n-n-n correlators as observed experimentally. The long-term algebraic decay is an emergent dissipative many-body phenomenon, where both the dissipation and strong interactions play a crucial role. This dynamics is independent of the details of the particular initial state, which are erased during the early part of the decay before the power law behavior is established. Experimentally, we observe that the time scale governing this early decay is much shorter than $\gsp^{-1}$. An accelerated early dynamics is qualitatively consistent with the mean-field numerical simulations of \cite{pichler2010a}, and intuitively expected in an interacting system. A recent publication\,\cite{pan2020a} suggests that the initial decay should be controlled by a ``non-Hermitian susceptibility'' that characterizes the response of the initial system to dissipation. In analogy with its equilibrium counterpart, the non-Hermitian susceptibility could potentially be useful as a novel probe of many-body properties, for instance of the momentum distribution. We leave a deeper investigation of this question for future work.

\begin{acknowledgements}
It is a pleasure to dedicate this article to Jean Dalibard. We are genuinely grateful to be a part of the research group that he built at LKB. Each of us owe a great deal to Jean, not only for his essential contributions in our joint projects, but also for the continuous inspiration and education. 
\end{acknowledgements}

\appendix

\section{Theory of spontaneous emission from a Bose-Hubbard quantum gas}
\label{app:theoryRelax}

In this Appendix, we discuss the quantum-optical theoretical framework that allows one to describe the behavior of an ultracold gas of interacting atoms driven by near-resonant light. This framework (sometimes called ``generalized optical Bloch equations'') generalizes the well-known optical Bloch equations\,\cite{cct_atomphoton} to account for the quantized motion of atoms (here, in a periodic lattice potential). Such a generalization has been studied in depth in the context of laser cooling\,\cite{dalibard1985a,castin1991a,cirac1992a}. The generic equations of motion take the form of a master equation for the atomic density operator $\{ \hat{\rho}_{\alpha,\beta} \}_{\{\alpha,\beta=e,g\}}$, where each component $\hat{\rho}_{\alpha,\beta}$ is an operator with respect to the motional degrees of freedom. Here our treatment also takes light-induced many-body effects into account following Refs.\,\cite{lehmberg1970a,lehmberg1970b,ellinger1994a,morice1995a,pichler2010a}.

\begin{figure*}[!ht]
	\centering
	\includegraphics[width=\textwidth]{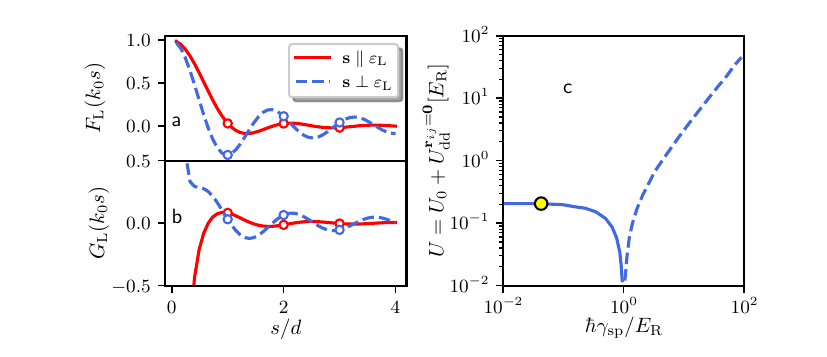}
	\caption{Real part (\textbf{a}) and imaginary part (\textbf{b}) of $K_{\mathrm{L}}(\bm{s})= F_{\mathrm{L}}(\bm{s})+\ii G_{\mathrm{L}}(\bm{s})= \epsilonL \cdot \bar{\bar{K}}(\bm{s}) \cdot \epsilonL^\ast$. The plots are drawn with our experimental parameters, $\lambda_{\mathrm{lat}}/\lambda_0 \approx 759/556 \approx 1.37$ and $\epsilonL=\bm{e}_x$ and for two orthogonal directions $\bm{s})$ along $\bm{e}_x$ and $\bm{e}_z$. \textbf{c}: on-site Hubbard interaction energy. The calculation includes the contribution of van der Waals interaction and of dipole-dipole interactions. The solid (respectively, dashed) curve corresponds to $U>0$ (resp., $U<0$). The yellow dot marks the experiments presented in the main text.}
	\label{fig8}
\end{figure*}

In the regime of weak laser excitation, saturation effects can be neglected and the \textit{internal} degrees of freedom can be eliminated adiabatically\,\cite{cct_atomphoton}. This approch leads to an effective Lindblad master equation (\ref{eq:lindblad}) describing the dynamics of the \textit{motional degrees of freedom} in the electronic ground state $g$. The Hamitonian is $\hat{H}=\hat{H}_{0}+\hat{V}_{\mathrm{dd}}$, with $\hat{H}_{0}$ the bare many-body Hamiltonian in the absence of the excitation laser, including interactions and the optical lattice potential, and with
\begin{align}\nonumber
\hat{V}_{\mathrm{dd}}  & =  \hbar \gsp \int d^3\bm{r}_1 d^3\bm{r}_2 \, G_{\mathrm{L}}(\bm{r}_{12})\ee^{-\ii \bmkL  \cdot \bm{r}_{12}}
\nonumber\\
& \hspace{2cm}
\big( \hat{\Psi}(\bm{r}_1)  \hat{\Psi}(\bm{r}_2) \big)^\dagger \hat{\Psi}(\bm{r}_1) \hat{\Psi}(\bm{r}_2)
\end{align}
the interaction  mediated by light scattering between the induced electric dipoles carried by each atom. We note $\hat{\Psi}(\bm{r}_\alpha)$ the field operator  annihilating an atom in the ground state at position $\bm{r}_\alpha$ and $\bm{r}_{12}=\bm{r}_{1}-\bm{r}_{2}$ the relative distance. The many-body dissipator in the Lindblad master equation  (\ref{eq:lindblad}) takes the form
\begin{align} 
 \mathcal{K}_{\mathrm{m-b}}\big[ \hat{\rho} \big]=&   \gsp   \int d^3\bm{r}_1 d^3\bm{r}_2 \, F_{\mathrm{L}}(\bm{r}_{12})\ee^{-\ii \bmkL  \cdot \bm{r}_{12}} \\
 \nonumber
 & \hspace{1cm}
 \left( \hat{n}(\bm{r}_1)  \hat{\rho} \hat{n}(\bm{r}_2) - \frac{1}{2} \Big\{ \hat{n}(\bm{r}_1) \hat{n}(\bm{r}_2) , \hat{\rho} \Big\} \right),
\end{align}
with $\hat{n}(\bm{r})= \hat{\Psi}^\dagger(\bm{r}) \hat{\Psi}(\bm{r})$ the particle density operator. The function $K_{\mathrm{L}}= F_{\mathrm{L}}+\ii G_{\mathrm{L}}= \epsilonL \cdot \bar{\bar{K}} \cdot \epsilonL^\ast$ is determined by the laser polarization $\epsilonL$ and by the so-called dyadic Green function $K_{m,m'}(\bm{r}_{12})$, the propagator of a photon from $\bm{r}_1$ to  $\bm{r}_2$ with polarization changing from axis $m$ to $m'$\,\cite{lehmberg1970a,lehmberg1970b}. The two functions $F_{\mathrm{L}}$ and $G_{\mathrm{L}}$ are shown in Fig.\,\ref{fig8}\textbf{a,b} for our particular experimental situation with $\lambda_{\mathrm{lat}}/\lambda_0 \approx 759/556 \approx 1.37$ and $\epsilonL=\bm{e}_x$.

We expand the master equation in the Wannier basis. For simplicity and to avoid notational overload, we assume that the Wannier functions $w(\bm{r}-\bm{r}_i)$ are tightly localized and use a tight-binding approximation. For one-body terms in the Hamiltonian, we only retain the dominant contributions involving on-site energies and nearest-neighbor tunnelling. The most generic form for two-body terms involves four sites labeled $i,j,k,l$. We only keep the dominant terms with $i=j$ and $k=l$. The bare many-body Hamiltonian $\hat{H}_0$ then reduces to the well-known Bose-Hubbard Hamiltonian,
\begin{align}
\hat{H}_{\mathrm{BH}}  & =-J_0 \sum_{\langle i,j \rangle} \hat{a}_{i}^\dagger \hat{a}_{j}
+
\frac{U_0}{2} \sum_{i} \hat{n}_{i}( \hat{n}_{i}-1),
\end{align}
with $J_0, U_0$ the tunnelling and interaction energies for atoms in the fundamental band and with $\hat{n}_{i}=\hat{a}_{i}^\dagger \hat{a}_{i}$ the on-site particle number operator. The d-d interactions and dissipator take the form
\begin{align}
\hat{V}_{\mathrm{dd}}   \simeq &  \sum_{i} U_{\mathrm{dd}}^{\bm{0}} \hat{n}_{i} (\hat{n}_{i}-1)  + \sum_{i\neq j} U_{\mathrm{dd}}^{\bm{r}_{ij}} \hat{n}_i \hat{n}_j,\\
\mathcal{K}_{\mathrm{m-b}}\big[ \hat{\rho} \big] & \approx  \sum_{ij} \Lambda_{\bm{r}_{ij}} \Big( \hat{n}_i \hat{\rho} \hat{n}_j - \frac{1}{2} \left\{ \hat{n}_i \hat{n}_j ,  \hat{\rho}\right\} \Big),
\end{align}
with matrix elements
\begin{align}
\frac{U_{\mathrm{dd}}^{\bm{r}_{ij}}}{\hbar\gsp} & =  \int d^3\bm{r}_1 d^3\bm{r}_2 \, G_{\mathrm{L}}(\bm{r}_{12}+\bm{r}_{ij})
\vert w(\bm{r}_1)\vert^2 \vert w(\bm{r}_2) \vert^2,\\
\label{eq:Lambdaij}
\frac{ \Lambda_{\bm{r}_{ij}}}{\gsp} &  =  \int d^3\bm{r}_1 d^3\bm{r}_2 \, F_{\mathrm{L}}(\bm{r}_{12}+\bm{r}_{ij}) \vert w(\bm{r}_1)\vert^2 \vert w(\bm{r}_2) \vert^2.
\end{align}

For $\bm{r}_{ij} \neq \bm{0}$, the functions vary slowly on the scale of the Wannier functions, and the d-d interaction energy can be approximated as
\begin{align}\label{eq:Udd_off}
U_{\mathrm{dd}}^{ \bm{r}_{ij} \neq \bm{0} } &  \simeq \hbar \gamma_{\textrm{sp}} G_{\mathrm{L}}(\bm{r}_{ij}).
\end{align}
Similarly, transition rates for any $\bm{r}_{ij}$ are approximately given by
\begin{align}
 \Lambda_{\bm{r}_{ij} } &  \simeq \gamma_{\textrm{sp}} F_{\mathrm{L}}(\bm{r}_{ij}).
\end{align}

In our experimental situation, the ratio $\lambda_{\mathrm{lat}} / \lambda_0$ is such that the function $F$ is small in all directions when $\bm{r}_{ij} \neq 0$. As a result, one can simplify the dissipative term and consider a ``zero-range model'' with
\begin{align}
 \Lambda_{\bm{r}_{ij} } &  \simeq \gamma_{\textrm{sp}} \delta_{ij}.
\end{align}
Off-site terms due to d-d interactions are similarly small and typically negligible even in comparison with the tunnelling energy. 

The on-site d-d interaction parameter $U_{\mathrm{dd}}^{\bm{0}}$ must be treated differently. For this term, the approximation leading to Eq.\,(\ref{eq:Udd_off}) for off-site terms is invalid, due to the $1/r^3$ divergence of d-d interactions at short distances\,\cite{lahaye2009a}. In three dimensions, this divergence is regulated by angular integration in three dimensions\,\cite{lahaye2009a}. Fig.\,\ref{fig8}\textbf{c} shows a numerical evaluation of the on-site interaction matrix element $U = U_0 + U_{\mathrm{dd}}^{ \bm{0}}$. For the parameters used in this work, the d-d term is a small correction to the total interaction energy and can be safely neglected. 

For higher dissipation rates comparable to or greater than $\ER/\hbar$, the dipole-dipole interaction energy can be substantial. In our experimental configuration, each site of the optical lattice resembles a pancake-shaped trap (the confinement is always stronger in the vertical direction), and the atomic electric dipoles are primarily along $(\bm{e}_x+\bm{e}_y)/ \sqrt{2}$. In this configuration, dipole-dipole interactions are attractive. As a result, the on-site interaction energy vanishes when the d-d interaction energy exactly cancels the background value, and becomes negative above. This feature could be used in future experiments to manipulate interactions among $^{174}$Yb atoms, or similar atoms lacking accessible Feshbach resonances.


\section{Decay of coherence for non-interacting atoms}
\label{app:1pdecay}

With the same approximations as in Appendix\,\ref{app:theoryRelax}, the dissipator describing the effect of spontaneous emission on the motional state of a single atom in the fundamental band of an optical lattice reads
\begin{align}
\mathcal{D}\big[\hat{\rho}\big] & =  \gsp \sum_{i,j} \Lambda_{x_{ij}} \left( \hat{n}_{i} \hat{\rho} \hat{n}_{j}
-\frac{1}{2} \{ \hat{n}_{i} \hat{n}_{j} , \hat{\rho}\} \right),
\end{align}
where $\Lambda_{x_{ij}}$ is defined in Eq.\,(\ref{eq:Lambdaij}) and where $x_{ij}=x_{j}-x_{j}$. The dissipator leads to an exact equation of motion for the spatially averaged coherence $C_s$,
\begin{align}
\frac{d C_s}{dt} & = - \gsp \big(1-\Lambda_{s}\big) C_s.
\end{align}
Therefore, $C_s$ decays exponentially with a rate constant $\gsp (1-\Lambda_{s})$ varying from zero when $s=0$ to $\approx \gsp$ when $ s $ is much larger than the light wavelength $\lambda_0$. If we take a ``zero range'' limit where $\lambda_0$ is much shorter than the lattice spacing, the rate constant becomes $\gsp (1-\delta_{s,0})$. This leads to Eq.\,(\ref{eq:Cs_independent}) in the main text. 

\bibliography{decoherenceBH_v2.bib}
\bibliographystyle{apsrev}

\end{document}